\documentclass[pdflatex,sn-mathphys-num]{sn-jnl}

\usepackage{graphicx}%
\usepackage{multirow}%
\usepackage{amsmath,amssymb,amsfonts}%
\usepackage{amsthm}%
\usepackage{mathrsfs}%
\usepackage[title]{appendix}%
\usepackage{xcolor}%
\usepackage{textcomp}%
\usepackage{manyfoot}%
\usepackage{booktabs}%
\usepackage{multirow}
\usepackage{multicol}
\usepackage{algorithm}%
\usepackage{algorithmicx}%
\usepackage{algpseudocode}%
\usepackage{listings}%
\usepackage{bm,bbm}
\usepackage{xr-hyper}
\usepackage{geometry}

\theoremstyle{thmstyleone}%
\theoremstyle{thmstyletwo}%

\theoremstyle{thmstylethree}%

\begin{document}

\title[VAE for Mutational Signatures]{VAE-MS: An Asymmetric Variational Autoencoder for Mutational Signature Extraction}

\author*[1, 2]{Ida Egendal}
\author[1, 2]{Rasmus Froberg Brøndum}
\author[3]{Dan J Woodcock}
\author[4]{Christopher Yau}
\author[1,2]{Martin Bøgsted}

\affil[1]{\orgdiv{Center for Clinical Data Science}, \orgname{Aalborg University and Aalborg University Hospital}, \orgaddress{\street{Aalborg}, \postcode{9260}, \country{Denmark}}}
\affil[2]{\orgdiv{Clinical Cancer Research Center}, \orgname{Aalborg University Hospital}, \orgaddress{\street{Aalborg}, \postcode{9000}, \country{Denmark}}}
\affil[3]{\orgdiv{Nuffield Department of Surgical Sciences}, \orgname{University of Oxford}, \orgaddress{\street{Oxford}, \postcode{OX3 9DU}, \country{England}}}
\affil[4]{\orgdiv{Nuffield Department of Women’s and Reproductive Health}, \orgname{University of Oxford}, \orgaddress{\street{Oxford}, \postcode{OX3 9DU}, \country{England}}}

\abstract{Mutational signature analysis has emerged as a powerful method for uncovering the underlying biological processes driving cancer development. However, the signature extraction process, typically performed using non-negative matrix factorization (NMF), often lacks reliability and clinical applicability. To address these limitations, several solutions have been introduced, including the use of neural networks to achieve more accurate estimates and probabilistic methods to better capture natural variation in the data. In this work, we introduce a Variational Autoencoder for Mutational Signatures (VAE-MS), a novel model that leverages both an asymmetric architecture and probabilistic methods for the extraction of mutational signatures. \newline
VAE-MS is compared to with three state-of-the-art models for mutational signature extraction: SigProfilerExtractor, the NMF-based gold standard; MUSE-XAE, an autoencoder that employs an asymmetric design without probabilistic components; and SigneR, a Bayesian NMF model, to illustrate the strength in combining a nonlinear extraction with a probabilistic model. In the ability to reconstruct input data and generalize to unseen data, models with probabilistic components (VAE-MS, SigneR) dramatically outperformed models without (SigProfilerExtractor, MUSE-XAE). The NMF-baed models (SigneR, SigProfilerExtractor) had the most accurate reconstructions in simulated data, while VAE-MS reconstructed more accurately on real cancer data. Upon evaluating the ability to extract signatures consistently, no model exhibited a clear advantage over the others.\newline
Software for VAE-MS is available at \href{https://github.com/CLINDA-AAU/VAE-MS}{GitHub}.
}

\keywords{Variational Autoencoders, Mutational Signatures}

\maketitle
\section{Introduction}
Mutational signature analysis is a subfield in genomics that focuses on identifying patterns of somatic mutations in cancer genomes and linking them to biological processes that cause the disease. Mutational signature analysis results in a matrix representing the signatures of mutagenic processes, i.e. mutational signatures, and an exposure matrix dictating the amount of mutations that can be attributed to specific processes in the analyzed cancer genome. 
Since the introduction of non-negative matrix factorization (NMF)-based mutational signature extraction by \citep{Alexandrov2013}, the biological relevance of mutational signatures has been well established through extensive validation, demonstrating their link to different biological processes \citep{Nemeth, etiologi, Drost2017}. Mutational signature analysis can reveal crucial information about the underlying biological mechanisms of the individual tumor and has the potential to inform and improve treatment decisions.\newline 

However, the clinical potential of mutational signatures is not yet fully utilized, although the field has been around for more than a decade \citep{hematological_malignancies}. A primary source of hesitation in using mutational signatures is the existence of overly specific and redundant mutational signatures. For example, the leading library of mutational signatures (COSMIC) introduced seven signatures associated with mismatch repair (MMR) in its third version \citep{COSMIC}, while \cite{Nemeth} demonstrated that a more concise set consisting of two mutational signatures could sufficiently describe MMR-related mutational processes. This redundancy likely stems from several fundamental limitations of the modeling techniques used to extract the signatures. One major cause is that the strictly linear nature of NMF is too simplistic to fully capture the complexity of mutational processes in cancer genomes. Specifically, two mutational signatures related to mutations in the proofreading domain of the polymerase epsilon (POLE) gene and the MMR pathway have, in fact, been shown to interact nonlinearly on the genome \citep{nonlin}. This simplified model specification can result in the artificial introduction of additional signatures to compensate for a poor fit, thus resulting in redundant and ill-fitting signatures. Another contributing factor is the overdispersion observed in mutational data \citep{SIGMOS}, which suggests unaccounted for variance in the model. A strictly deterministic approach, such as NMF, struggles to model this inherent heterogeneity, and additional signatures may be introduced to absorb variability that the model fails to explain, inflating the number of extracted signatures. Lastly, NMF suffers from inherent non-uniqueness issues, which means that multiple equally valid decompositions can lead to identical reconstructions \citep{Laursen2022}, affecting the reliability and consistency of signature identification. Together, these limitations contribute to the generation of redundant and overly specific mutational signatures, ultimately reducing the clinical utility of mutational signature analysis.\newline

To address these limitations, recent studies have explored alternative approaches to extract mutational signatures, including deep learning-based methods such as autoencoders \citep{MUSEXAE, Pei2020, mig}. In particular, \cite{MUSEXAE} proposed an asymmetric autoencoder architecture with a deep nonlinear encoding network to capture complex patterns while maintaining interpretability through a linear decoding network. In parallel, Bayesian NMF, a probabilistic extension of NMF, uses stochastic factor matrices to improve robustness to data variability \citep{signeR, sigfit, zito}, and a review of computational tools for mutational signature extraction indeed found that probabilistic methods generally outperformed purely NMF-based methods \citep{Omichessan2019}. However, existing methods for extracting mutational signatures retain the limitations of linear and/or deterministic modeling.\newline

In this study, we extend the asymmetric autoencoder framework introduced by \cite{MUSEXAE} by incorporating probabilistic modeling in the form of the first variational autoencoder for mutational signature extraction, VAE-MS, which encodes the input into a latent Poisson distribution and does not assume linearity, offering greater flexibility in capturing different and more complex mutational patterns. A Poisson distributed latent space is assumed to accommodate the non-negative nature of the exposure matrix while remaining on the original scale of the data. We hypothesize that integrating nonlinearity with a probabilistic latent space will yield superior performance compared to existing approaches. To evaluate VAE-MS, we compared its performance with SigProfilerExtractor, an NMF-based approach \citep{SigProfilerExtractor}, MUSE-XAE, and SigneR, a Bayesian NMF-based approach \cite{signeR}.

\section{Materials and Methods}
\begin{figure}
    \centering
    \includegraphics[width=\linewidth]{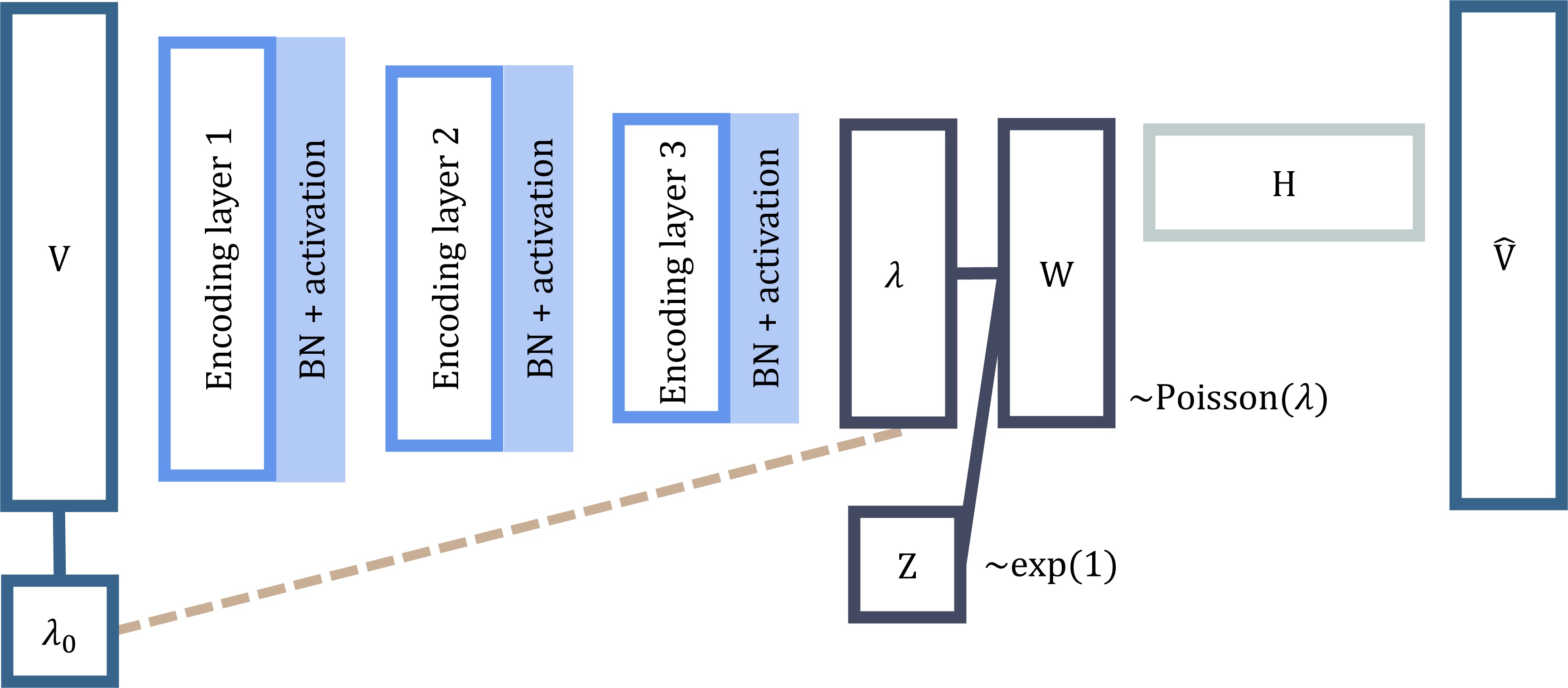}
    \caption{Schematic representation of VAE-MS. BN: batch normalization.}
    \label{fig:VAE-MS}
\end{figure}
\subsection{The Variational Autoencoder Mutational Signature (VAE-MS) Model} \label{sec:VAE-MS}
VAE-MS consists of a deep encoding network with layers of decreasing dimensionality, encoding the input data into a rate parameter of the exposure matrix. The exposure matrix is drawn from a Poisson distribution and is subsequently multiplied with a non-negative matrix to reconstruct the input data. This decoding matrix represents the signature matrix and constitutes the entire decoding network. A graphical representation of the architecture of VAE-MS can be seen in Figure \ref{fig:VAE-MS}.\newline
\textbf{Input Data:} The input is a normalized mutation matrix $V \in \mathbb{Z}_+^{(N\times M)}$, where each row represents an individual patient ($1, \dots, N$), and each column corresponds to a mutation type ($1, \dots, M$). \newline
\textbf{Encoding Network:} The encoding network, consists of three fully connected layers, each reducing the dimensionality of the input. Every layer is followed by batch normalization and an activation function. The final encoded representation is then transformed by a fully connected layer to generate the matrix of rate parameters, $\lambda \in \mathbb{R}^{N\times K}$, of the latent distribution.\newline
\textbf{Latent Representation:} Given the learned matrix of rate parameters, $\lambda$, the latent distribution is assumed to be:
\begin{equation}
    W_{n,k} \sim \text{Poisson}(\lambda_{n,k}), \quad \text{for } n=1,\dots,N \text{ and } k=1,\dots,K.
\end{equation}
 The Poisson-distributed values are drawn using the novel Poisson reparameterization trick where $W_{n,k}$ is expressed as a function of an infinite series $Z_{n,k}=\{Z_{n,k,1},Z_{n,k,2},\dots\}$ of i.i.d. $\text{Exponential(1)}$-distributed values and the parameter $\lambda_{n,k}$, by counting the number of occurrences in $[0,1]$ of a Poisson process defined by the waiting times $\lambda_{n,k}\cdot Z_{n,k,1}, \lambda_{n,k}\cdot  Z_{n,k,2}, \dots$ \citep{PVAE}.\newline
\textbf{Decoding Network:} The decoding network reconstructs the mutational count matrix using a single linear transformation without bias terms:
\begin{equation}
    \hat{V} = WH,
\end{equation}
where $W$ represents the exposure matrix and $H$ represents the mutational signature matrix. This structure ensures interpretability by maintaining similarity to traditional NMF-based decomposition methods \citep{Pei2020, MUSEXAE, mig}.\newline
\textbf{Scaling:} In mutational signature extraction, the matrices $W$ and $H$ are typically scaled so that the sum of each row of $H$ equals one, while the rows sums of $W$ estimate the row sums of the input matrix $V$. VAE-MS incorporates scaling directly in the forward pass, ensuring that training and computation of the KL-divergence occurs on the intended scale.

\subsubsection{Training and Optimization}
The training and optimization framework of VAE-MS is further elaborated and illustrated in Supplementary Materials Section 2 and Figure S1.\newline
\textbf{Normalization:} The input matrix is normalized to reduce the influence of hypermutated patients on the trained model. For patients with a total number of mutations that exceeds 100 times the number of features $(\sum_{m=1}^M v_{n,m} > 100\cdot M)$, each entry was scaled down by a factor $100\cdot M/\sum_{m=1}^M v_{n,m} $. This technique is known as 100X normalization \citep{100x, SigProfilerExtractor}. However, all the metrics reported in the present study are calculated using unnormalized data.\newline
\textbf{Prior Distribution:} We assume that the prior is a Poisson distribution with the matrix of prior Poisson rates, $\lambda_0 \in \mathbb{R}_+^{N\times K}$, initialized using NMF:
    \begin{align}
    \begin{split}
        V &\overset{\text{NMF}}{\approx} W_{\text{NMF}}\cdot H_{\text{NMF}},\\
        \lambda_0 &= W_{\text{NMF}} \cdot \text{diag}\left(\left\{ \sum_{m = 1}^M H_{\text{NMF } {i,k}}\right \}_{k = 1, \dots,K}\right).
        \end{split}
    \end{align}
Thus, the prior Poisson rates, $\lambda_0$, are the weight matrix of an NMF decomposition of $V$, scaled such that the rows of the basis matrix $H_{\text{NMF}}$ sum to one, as described in Section \ref{sec:VAE-MS} (Scaling). This choice of prior implies that NMF provides a reasonable starting point that can be further refined using VAE-MS. \newline
\textbf{Loss function:} To train the model, a Poisson likelihood is assumed, i.e. $v_n|w_n,\theta\sim\text{Poisson}(\sum_{k=1}^K W_{n,k}H_{k,m})$. Posterior inference is performed by maximizing a lower bound on the marginal log-likelihood, the evidence lower bound (ELBO):
\begin{equation}\label{eq:ELBO}
    \mathcal{L}_{\theta,\phi} = \sum_{n=1}^{N}\mathbb{E}_{q_\phi(w_n|v_n)}\left[\log p_\theta(v_n|w_n)\right] -\mathcal{D}_{\text{KL}}(q_\phi (w_n|v_n)|| p(w)),
\end{equation}
where $p_\theta$ is the likelihood of the reconstructed data $\hat{V}$, $q_\phi$ is the approximation of the posterior distribution of $W$, and $p$ denotes the prior distribution of $W$. To control the trade-off between reconstruction accuracy and latent space regularization the ELBO is altered to have the KL divergence term weighted by a hyperparameter $\beta>0$ \citep{betavae}:
\begin{align}\label{eq:B-ELBO}
\begin{split}
    \mathcal{L}_{\theta,\phi,\beta} =& \sum_{n=1}^{N}\mathbb{E}_{q_\phi(w_n|v_n)}\left[\log p_\theta(v_n|w_n)\right] \\
    -&\beta\mathcal{D}_{\text{KL}}(q_\phi (w_n|v_n)|| p(w)).
     \end{split}
\end{align}
\textbf{Data Splits:} All data sets are split into 80/20 training and test splits. Subsequently, the training data is further split into a smaller training set and a validation set using an additional 80/20 split, resulting in an overall 64/16/20 split for training, validation, and test. The validation set is used to guide the hyperparameter optimization as well as the early stopping mechanism of VAE-MS. Once training is completed, the final training loss is calculated by performing a forward pass of the pooled, unnormalized training and validation data through the trained model. \newline
\textbf{Hyperparameter Optimization:} To determine the optimal configuration of the hyperparameters, a Bayesian optimization-based sweep is performed across a number of candidate parameter sets given a fixed signature count \citep{BayesOpt}. The sweeps are governed by and tracked using Weights and Biases \citep{wandb} and are configured to minimize the validation loss. The hyperparameter search domains are listed in Supplementary Table S1.\newline
\textbf{Early Stopping:} An early stopping criterion is applied, with training stopping at 2000 epochs or if the validation loss does not improve over 50 epochs. The model at the epoch with the lowest validation loss is selected.\newline
\textbf{Stability Runs:} To assess the stability of the solutions from a sweep of a given split and the number of signatures ($k$), the hyperparameter configuration that produces the lowest validation loss is used to train ten additional models. \newline 
\textbf{Number of Signatures:} To determine the optimal number of mutational signatures for a split, the average validation loss and silhouette score (Supplementary Materials Section 3.5) of the signatures extracted in the stability runs for each candidate signature count ($k = k_{\min}, \dots, k_{\max}$) are calculated. The optimal number of signatures is identified as the point where the difference between the silhouette score and the average validation loss is maximized, following the methods of \citep{SigProfilerExtractor, MUSEXAE}.

\subsection{External Resources}
\subsubsection{Data} 
Evaluation is performed on both simulated and cancer genomic data. All analyses in the present study are performed on the frequency distribution of single-base substitutions (SBS) and the nucleotides immediately to the left and right, leading to $4\cdot 6\cdot 4 = 96$ combinations, commonly denoted as the SBS96 representation. \newline
\textbf{PCAWG:} The Pan Cancer Analysis of Whole Genomes (PCAWG) Consortium is an international collaboration in which tumor-normal whole-genome sequencing (WGS) of 2780 mutational profiles from 38 different cancer types were performed \citep{PCAWG}. \newline
\textbf{Simulated Data:} The simulated data sets considered in the present study are Scenario 8 (S8) and Scenario 14 (S14) provided by \cite{SigProfilerExtractor}. S8 consists of 1000 simulated mutational profiles that emulate a mixture of renal cell carcinomas and ovarian adenocarcinomas. There are three ground-truth signatures in S8, which resemble the COSMIC signatures SBS3, SBS5, and SBS40. S14 is made to resemble PCAWG and contains 2700 simulated mutational profiles that emulate one of nine different cancer types (bladder transitional cell carcinoma, esophageal adenocarcinoma, breast adenocarcinoma, lung squamous cell carcinoma, renal cell carcinoma, ovarian adenocarcinoma, osteosarcoma, cervical adenocarcinoma, and stomach adenocarcinoma). S14 has 21 ground-truth signatures. For both S8 and S14 the simulations are derived from the matrix product between the ground-truth signatures and the corresponding exposures.
\subsubsection{Comparison Models}
To assess the performance of VAE-MS, it is compared against three state-of-the-art methods:
\begin{itemize}
    \item SigProfilerExtractor \citep{SigProfilerExtractor} – The gold standard for mutational signature extraction.
    \item MUSE-XAE \citep{MUSEXAE} – A deep learning-based asymmetric autoencoder.
    \item SigneR \citep{signeR} – A probabilistic empirical Bayes model.
\end{itemize}
All models are elaborated in Supplementary Materials Section 1. As no tuning is needed on the comparison models, training is performed on the pooled training and validation data.
\subsection{Performance Measures}
Reconstruction accuracy is measured using both the Kullback-Leibler Divergence (KLD - Supplementary Materials Equation S1) and the mean-squared error (MSE - Supplementary Materials Equation S2). VAE-MS, SigProfilerExtractor, and MUSE-XAE are optimized with respect to KLD, whereas SigneR is optimized with respect to MSE.\newline

Signature similarity is measured using the average cosine similarity (ACS) (Supplementary Materials Section 3.3), where the signatures are aligned by the Hungarian algorithm (Supplementary Materials Section 3.2). To evaluate the stability and consistency of the estimated signatures and their labeling, Pairwise Average Cosine Similarity (PACS) and Silhouette Score (SS) are used (Supplementary Materials Sections 3.4 and 3.5). PACS measures the ACS of all pairwise combinations of signature sets estimated by a given model between the data splits and is a measure of the stability of the estimated signatures. SS measures the consistency and separability of the signature labeling between the data splits and is therefore only reported when the number of signatures is fixed across the 10 splits.\newline

For the simulated data, the fraction of true exposures contained in the $95\%$ credibility intervals (Supplementary Materials Section 3.6) was reported for the models where it was available, i.e. VAE-MS and SigneR. 
\begin{figure*}
    \centering
     \includegraphics[width=\linewidth]{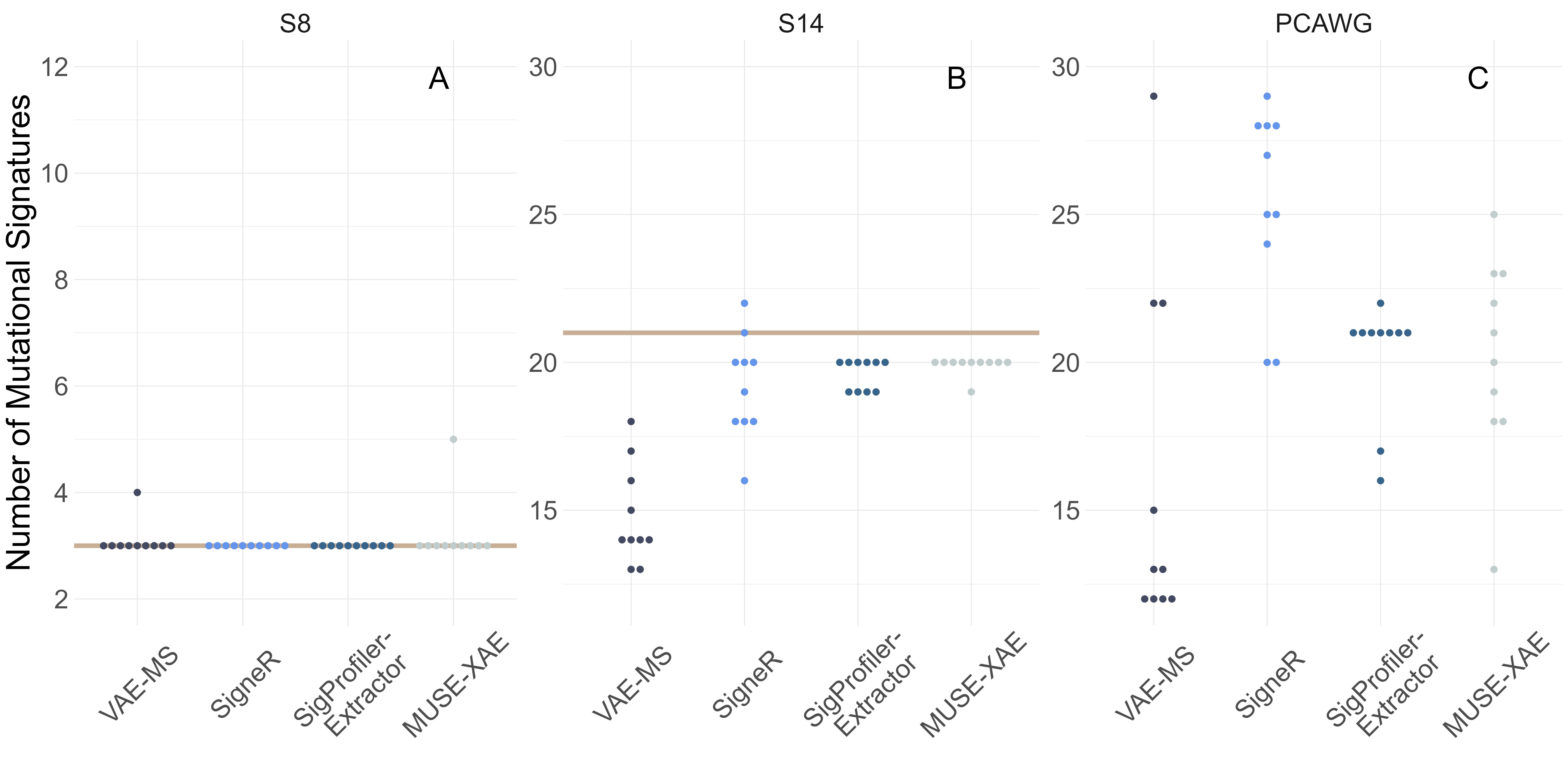}
    \caption{Beeswarm plots of the number of signatures suggested by each model for extraction in S8 (A), S14 (B), and PCAWG (C) over the 10 splits by all compared models. The true number of signatures in the simulated datasets is highlighted with a tan line.}
    \label{fig:nsigs}
\end{figure*}
\section{Results}
Each data set (S8, S14, and PCAWG) was split into 10 training/validation/test sets.
\subsection{Simulated data}
The optimal number of signatures was searched over $k = 2,\dots,12$ for S8 and $k = 12,\dots, 30$ for S14, and the average validation loss and silhouette score against the number of signatures can be seen in Figures S2 and S3. The number of signatures selected by each model is depicted in Figure \ref{fig:nsigs} (A + B). The NMF-based methods (SigProfilerExtractor, SigneR) recovered the true number of signatures in all 10 splits of S8. In contrast, the autoencoder-based methods (MUSE-XAE, VAE-MS) recovered the true number in 9 of 10 splits. For S14, the number of signatures selected by VAE-MS ranged from 13 to 18; therefore, the true number was not identified in any split. SigProfilerExtractor and MUSE-XAE did not identify the true number of signatures in any split, but consistently selected either 19 or 20 signatures. SigneR identified the true number of signatures exactly once, with values ranging from 16 to 22.\newline

\begin{table*}[t]
\tabcolsep=0pt
\caption{The training and test performance of all models on the simulated data sets S8 and S14 with the number of signatures fixed at the true number of signatures. $\mathcal{D}_{\text{KL}}$: Kullback-Leibler divergence, $\mathcal{D}_\text{MSE}$: Mean-square error, ACS: Average Cosine Similarity to true signature set, PACS: Pairwise Average Cosine Similarity, SS: Silhouette Score, SigProEx: SigProfilerExtractor. Lowest error ($\mathcal{D}_{\text{KL}}, \mathcal{D}_{\text{MSE}}$), highest similarity measures (ACS, PACS, and SS) and highest fraction of exposures contained in CI are highlighted with \textbf{bold text}.}
\label{tab:sim_perf}
\begin{tabular*}{\textwidth}{@{\extracolsep{\fill}}ll|rrrrrrrrr@{\extracolsep{\fill}}}
                     &                                           & \multicolumn{2}{c}{$\mathcal{D}_{\text{KL}}$} & \multicolumn{2}{c}{$\mathcal{D}_\text{MSE}$} &   \multicolumn{3}{c}{Signature}  &  \multicolumn{2}{c}{Exposures in CI} \\ 
Data              & Model                                     & Train         & Test        & Train        & Test        & ACS  & PACS  & SS & Train  & Test\\ \hline
\multirow{4}{*}{S8}  & VAE-MS &    $13.1$      &     $5.9$   &    $407$   &    $421$        &  $0.96$   &   $\bm{0.99}$& $0.98$ & $8.4\%$ &  $7.3\%$  \\
                     & SigneR      &     $7.9$    &    $\bm{3.1}$    &    $2$        &      $\bm{5}$      &   $\bm{0.99}$  & $\bm{0.99}$ & $\bm{0.99}$ & $\bm{36\%}$    &  $\bm{24\%}$  \\
                     & SigProEx  &     $\bm{3.1}$   &    $3.2$  &   $\bm{0.12}$       &   $12$         &  $\bm{0.99}$   &  $\bm{0.99}$& $\bm{0.99}$ & - &  -  \\
                     & MUSE-XAE             &      $4.2$  &    $4.2$  &   $458$       &   $465$         &  $\bm{0.99}$   &      $\bm{0.99}$    &  $\bm{0.99}$ & -  &  - \\ \hline
\multirow{4}{*}{S14} & VAE-MS  &    $9.1$       &   $5.2$         &    $2231$         &      $2347$      &   $0.80$  &   $\bm{0.99}$ & $0.48$  & $\bm{24.4\%}$  & $\bm{24.6\%}$ \\
                     & SigneR       &    $\bm{3.0}$          &    $\bm{3.0}$        &       $\bm{21}$      &   $\bm{33}$         &  $\bm{0.95}$   &    $0.98$  & $0.83$ & $1.3\%$  & $0.8\%$ \\
                     & SigProEx  &      $\bm{3.0}$        &     $14.4$       &     $144$        &     $19831 $       &   $0.94$ &  $0.81$ & $\bm{0.96}$ &   - & - \\
                     & MUSE-XAE             &     $16.4$         &    $15.9$        &      $ 184427$       &      $97669$      &  $0.94$   &  $0.98$  &  $0.91$   &  - & -
\end{tabular*}
\end{table*}
Table \ref{tab:sim_perf} shows that the NMF-based models (SigneR, SigProfilerExtractor) achieved a higher reconstruction accuracy in both training and test data across both simulated datasets, as measured by the KLD ($\mathcal{D}_{\text{KL}}$) and the mean-squared error ($\mathcal{D}_\text{MSE}$).  Specifically, SigProfilerExtractor had the lowest training losses, while SigneR achieved the lowest test losses for S8. For S14, SigneR outperformed all other models across both metrics on both training and test data.\newline

VAE-MS had the lowest ACS with the true signature matrix, the difference being marginal in S8 ($0.96$ for VAE-MS versus $0.99$ for the remaining models) and more expressed in S14 ($0.80$ for VAE-MS versus $0.94$ for SigProfilerExtractor and MUSE-XAE and $0.95$ for SigneR). All models achieved a similar and high PACS $(\geq 0.98)$ between the ten splits of S8 and S14, except SigProfilerExtractor with a PACS of $0.81$ for S14. For S8, all models showed high SS $(\geq 0.98)$. For S14 the deterministic models showed higher SS (SigProfilerExtractor: $0.96$, MUSE-XAE: $0.91$) compared to SigneR $(0.81)$ and VAE-MS $(0.48)$.\newline

\begin{figure*}[!t]
    \centering
    \includegraphics[width =  \linewidth]{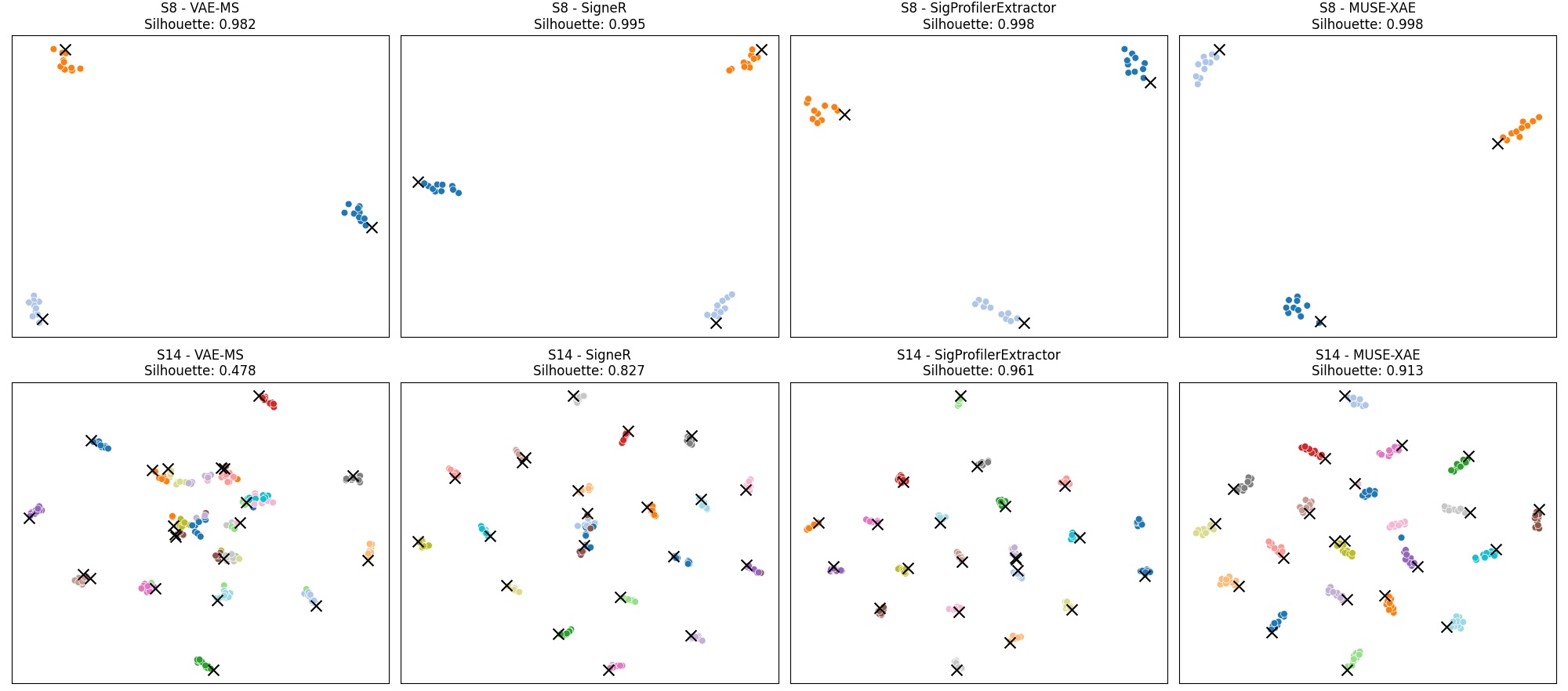}
    \caption{2D t-SNE embeddings of the estimated signatures across the 10 splits of the simulated data sets, with the number of signatures fixed to the true value. Embeddings of the true signatures are indicated with black 'x' markers.}
    \label{fig:tsne}
\end{figure*}

Figure \ref{fig:tsne} and Supplementary Figure S3 show 2D t-SNE \citep{tsne} embeddings of the signatures estimated by each model in the case where the number of signatures is fixed to the true number and chosen freely, respectively. The embeddings echo the patterns observed in Table \ref{tab:sim_perf}: The signatures of the deterministic models have high overlap with the true signatures (high ACS) and well-defined and separated clusters (high SS), while the probabilistic models have less distinct clusters (low SS).\newline

\begin{figure}
    \centering
    \includegraphics[width = \linewidth]{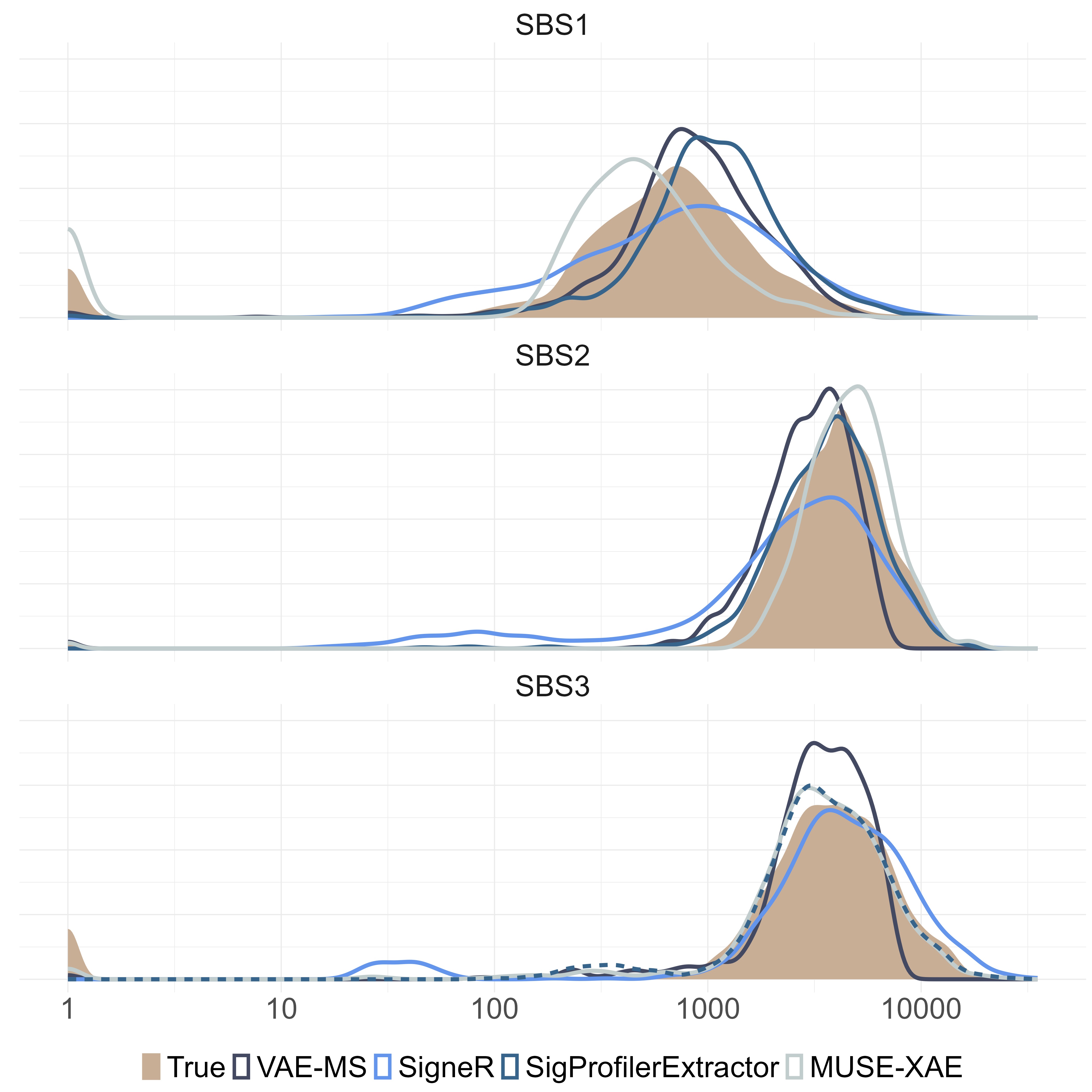}
    \caption{Density plots of the true and estimated exposures for the corresponding signature for the second split of the S8 data.}
    \label{fig:violin_exposures}
\end{figure}

Figure \ref{fig:violin_exposures} compares the density of the estimated exposures with the true exposures from the second split of the S8 data set. In this split, SigneR's $95\%$ credibility intervals contained $32.4\%$ of the true training exposures and $23.2\%$ of the true test exposures. For VAE-MS, the intervals contained $8.5\%$ and $9.1\%$ of the true training and test exposures, respectively. Table \ref{tab:sim_perf} reports the average fraction of true exposure values contained in the $95\%$ credibility intervals. Although neither model covered even half of the exposures, SigneR consistently covered a larger fraction in S8, whereas the opposite was observed in S14.\newline

A comparison between the results in Table \ref{tab:sim_perf}, where the number of signatures was fixed to the true value, and Table \ref{tab:PCAWG_perf}, where the number was freely selected, reveals the impact of constraining the models. For S8, VAE-MS and MUSE-XAE had lower average reconstruction losses when allowed to select the number of signatures freely. SigneR and SigProfilerExtractor maintain identical performance across both settings, as they identified the true number of signatures in all splits of S8. For S14, VAE-MS and MUSE-XAE again demonstrated reduced average reconstruction losses in the unconstrained setting. For SigneR, MSE losses increased drastically and for SigProfilerExtractor, training losses increased slightly, while the test losses increased drastically. Substantial changes in PACS were not observed, except for SigProfilerExtractor on S14, where PACS increased from $0.81$ to $0.99$.

\subsection{PCAWG}
Similarly to S14, the number of signatures for PCAWG was searched over $k = 12,\dots, 30$\, and the average validation loss and silhouette score against the number of signatures can be seen in Figure S4. The numbers selected by each model are shown in Figure \ref{fig:nsigs} (C). VAE-MS selected a lower number of signatures with high variability (mean: $16.2$, standard deviation: $5.68$). SigneR selected a higher number with a lower variability (mean: $25.4$, standard deviation: $3.10$). Both SigProfilerExtractor and MUSE-XAE selected a mid-range number (mean: $20.2$), but with different variability (standard deviation: $1.88$ and $3.35$, respectively).\newline

Table \ref{tab:PCAWG_perf} reports the average reconstruction errors and PACS between the 10 data splits. The probabilistic models (VAE-MS and SigneR) outperformed the deterministic models (SigProfilerExtractor and MUSE-XAE) in reconstruction accuracy, and VAE-MS yielded the most accurate reconstructions overall, with the lowest reconstruction errors in three of four metrics (training KLD, and training and test MSE). SigProfilerExtractor and MUSE-XAE achieved high PACS $(0.99)$, while VAE-MS and SigneR showed slightly lower PACS $(0.93)$.
\begin{table}[]
\centering
\caption{Training and test performance of all models on the simulated and PCAWG data with the number of signatures selected by the models. $\mathcal{D}_{\text{KL}}$: Kullback-Leibler divergence, $\mathcal{D}_\text{MSE}$: Mean-squared error, PACS: Pairwise Average Cosine Similarity. Lowest error ($\mathcal{D}_{\text{KL}}, \mathcal{D}_{\text{MSE}}$), highest PACS are highlighted with \textbf{bold text}.}
\label{tab:PCAWG_perf}
\begin{tabular}{ll|rrrrr}
                     &              &                      \multicolumn{2}{c}{$\mathcal{D}_{\text{KL}}$} & \multicolumn{2}{c}{$\mathcal{D}_\text{MSE}$}   \\
& Model                                     & Train         & Test        & Train        & Test     &  PACS  \\ \hline
\multirow{4}{*}{\rotatebox{90}{S8}} &VAE-MS &    $ 11.7$      &     $4.9 $   &    $255 $   &    $255 $          &   $\bm{0.99}$    \\
&SigneR      &     $7.9$    &    $ 3.1 $    &    $ 2 $        &      $ 5 $        & $\bm{0.99}$   \\
&SigProfilerExtractor &     $\bm{3.1}$   &    $3.2$  &   $\bm{0.12}$       &   $12$        &   $\bm{0.99}$ \\
&MUSE-XAE             &      $\bm{3.1}$  &    $\bm{3.1}$  &   $3.6$       &   $\bm{1.6}$            &     $\bm{0.99}$    \\ \hline
\multirow{4}{*}{\rotatebox{90}{S14}} &VAE-MS &    $4.7$      &     $4.5$   &    $1817$   &    $1861$          &   $0.96$    \\
&SigneR      &     $\bm{3.1}$&    $\bm{3.2}$    &    $\bm{215}$        &      $\bm{407}$        & $ 0.97 $   \\
&SigProfilerExtractor &      $3.2$   &    $28.8$  &   $319$       &   $728112$            &   $\bm{0.99}$ \\
&MUSE-XAE             &      $4.5$  &    $4.4$  &   $7204$       &   $2169$            &     $0.96$    \\ \hline
\multirow{4}{*}{\rotatebox{90}{PCAWG}} &VAE-MS &    $5.5 $      &     $\bm{5.2}$   &    $\bm{1876}$   &    $\bm{1718}$          &   $0.94$    \\
&SigneR      &     $\bm{4.2}$    &    $ 5.3 $    &    $ 5713 $        &      $ 53014 $        & $ 0.93 $   \\
&SigProfilerExtractor &    $10.8$ &  $19.4$  &  $166035$  &  $464576$      &   $\bm{0.99}$ \\
&MUSE-XAE             &    $11.4$      &  $10.7$    & $282514$      &       $216085$     &     $\bm{0.99}$    \\ \hline
\end{tabular}
\end{table}

\section{Discussion}
The results showed that the probabilistic models (VAE-MS, SigneR) outperformed the deterministic models (SigProfiler-Extractor, MUSE-XAE) in terms of reconstruction accuracy for PCAWG, whereas the NMF-based methods (SigneR, SigProfilerExtractor) in general reconstructed more accurately on the simulated data sets. All models displayed a high PACS ($>0.9$) in all data sets, except SigProfilerExtractor on S14 with a PACS of $0.81$. The deterministic models displayed high silhouette scores ($>0.9$) in both simulated data sets, where SigneR and VAE-MS displayed lower silhouette scores in S14 ($0.83$ and $0.48$, respectively).\newline

 The two simulated data sets used in the present study were derived from a matrix product of the true signatures and exposures, a process that inherently aligns with NMF-based models such as SigProfilerExtractor and SigneR. Consequently, these data sets may not be ideal for comparing the performance of nonlinear models such as VAE-MS and MUSE-XAE with NMF-based models, which indeed performed favorably in reconstructing the simulated data. The discrepancy between the simulated data's linear generating process and the nonlinear extraction methods indicates that VAE-MS and MUSE-XAE will not necessarily estimate the true factor matrices, but instead identify an alternative, possibly reduced, set of signatures. This tendency becomes evident in the low ACS between the VAE-MS and true signatures, and is a key limitation of VAE-MS: its difficulty in estimating the true number of mutational signatures in simulated settings, particularly in S14, where it repeatedly selects fewer signatures than expected. Furthermore, both VAE-MS and MUSE-XAE improve in all reconstruction measures of the simulated data when the number of signatures is unconstrained, indicating that these may identify a reduced, alternative set of signatures. \newline

The generally low fractions of true exposures contained in the credibility intervals generated by VAE-MS for the simulated data likely have multiple causes. Modeling the latent variable using a Poisson distribution appears to be a suitable choice given the count nature of the exposure matrix. However, mutational count data has been shown to be overdispersed when modeled with a Poisson distribution, and a Negative Binomial distribution may be a better fit \citep{SIGMOS, NBVAE}. Furthermore, choosing the Poisson distribution to model exposure counts around $0$ becomes critical as the rate of a Poisson distribution cannot assume the value $0$, although sufficiently low rates can generate realizations of $0$. Thus, the calculation of the credibility intervals of the VAE-MS exposures was based on Monte Carlo sampling instead of analytical calculations, and this sampling-based approach may affect the reliability and accuracy of these credibility intervals. Secondly, variational distributions are known to underestimate variation \citep{VIforStats}. These effects indicate that there is variance in the system that remains unaccounted for, causing narrower credibility intervals, and thus a lower fraction of true exposures contained within them.\newline

Examining only 10 different hyperparameter configurations when tuning VAE-MS is a relatively low amount, but was chosen to ensure a fair comparison across models. For future applications of VAE-MS, testing a larger number of hyperparameter configurations is recommended.

\section{Conclusion}
In this study, we introduced VAE-MS, the first variational autoencoder for the extraction of mutational signatures. We demonstrated that it outperforms three state-of-the-art models in reconstruction accuracy on real cancer genomics data. VAE-MS signatures display high stability between data splits, but low silhouette scores are indicative of inconsistent signature assignments across the splits.\newline
This study demonstrates how the combination of deep neural networks and probabilistic modeling enables more flexible and nonlinear extraction of mutational signatures, leading to superior reconstruction accuracy and potentially higher clinical utility.

\section{Funding}
\noindent This work was supported by the Danish Data Science Academy [DDSA-PhD-2022-005 to I.E.], which is funded by the Novo Nordisk Foundation [NNF21SA0069429] and VILLUM FONDEN [40516] and the Research Hive "REPAIR", which is funded by Aalborg University Hospital.

\section{Conflicts of Interest}
\noindent None to declare.

\section{Code- and Data Availability}
Source code for VAE-MS can be found at \href{https://github.com/CLINDA-AAU/VAE-MS}{GitHub}. Simulated data is available through \cite{SigProfilerExtractor} and PCAWG cancer data can be applied for access to at the \href{https://daco.icgc-argo.org/}{ICGC DACO Data Portal}.

\bibliography{library}



\end{document}